\documentclass[12pt]{article}
\usepackage{array,epsfig,amssymb}

\newcommand{\kms}{\mathrm{\;km\;s^{-1}}}

\begin{document}

\title{\large{\vskip -0.2 in Measuring the Growth of Structure with Spectroscopically
Identified Galaxy Groups and Clusters}}

\author{\normalsize{Marc Davis, Brian F. Gerke, Alison L. Coil,
Michael C. Cooper, 
Renbin Yan}\\
\emph{\normalsize{University of California--Berkeley}}\vspace{0.1in}\\
\normalsize{Jeffrey A. Newman}\\
\emph{\normalsize{Lawrence Berkeley National Laboratory}}\vspace{0.1in}\\
\normalsize{S. M. Faber, David Koo, Puragra Guhathakurta}\\
\emph{\normalsize{University of California--Santa Cruz}}}

\begin{titlepage}
\maketitle

\vskip -0.3in
Number counts of galaxy clusters offer a very promising 
probe of the Dark Energy (DE) equation-of-state parameter, $w$. The
basic goal is to measure abundances of 
these  objects as a function of redshift, compare this to a
theoretical prediction, and infer the values of cosmological
parameters. Various teams have proposed such a measurement,
including the South Pole Telescope, the Dark Energy Survey and  the
Red-Sequence Cluster Survey. The specific study discussed here detects
clusters and smaller galaxy groups in the 
three-dimensional distribution of galaxies inferred from a large
spectroscopic redshift survey.  This method allows the abundance, $N$,
of groups and clusters to be measured as a function of \emph{velocity
dispersion}, as well as of redshift, permitting a more sensitive test
of cosmology.


This test is one of the principal science goals of the
DEEP2 Galaxy Redshift Survey, a spectroscopic survey of $\sim 50000$
galaxies 
over a primary redshift range of $0.7\le z \le 1.4$, using the DEIMOS
spectrograph on the ten-meter Keck II telescope.  The survey, which is
now nearly complete, has surveyed $\sim 3$ square degrees on the sky to
a limiting magnitude of $R_{AB}=24.1$, with a sampling rate of $\sim
60\%$ in the targeted redshift range.  The full survey required 80
nights of observation at Keck.   

In addition, a multiwavelength suite of observations
is ongoing in a subregion of the DEEP2 area, the Extended Groth
Strip (RA: 14 17, Dec: +52 30).  Possible 
systematic errors in both DEEP2 and other cluster samples could
be controlled by comparing to these data. Observations in this region with
potential application to clusters include \emph{Chandra}
X-ray observations, infrared photometry with \emph{Spitzer} and
ground-based telescopes for stellar masses, optical space-
(\emph{HST}) and ground-based (CFHT Legacy Survey) imaging for weak
lensing studies, many-band photometric redshifts, and SZ observations.  

We discuss here the DE
constraints expected from DEEP2, as well as the projected constraints
for a similar survey with $\sim 20$ times as much sky coverage.  DEEP2
has the power to constrain $w$ to $20\%$ (1$\sigma$) without combining with other dark energy constraints; the larger
survey could constrain $w$ to $\sim 5\%$.

\end{titlepage}

\setcounter{page}{2}
\section{The Measurement}

By counting the abundance $N$ of groups and clusters as a function of
their redshift $z$ and velocity dispersion $\sigma$, we can probe the
dark energy in two ways.  The measured quantity in this test is 
\[
\frac{dN}{d\sigma dz} = \frac{dV}{dz}\frac{dn}{d\sigma}(\sigma,z).
\]
The comoving volume element, $dV/dz$, has a strong
dependence on $w$, as does the comoving
number density of dark matter halos, $n(\sigma,z)$.  The former
dependence can be written down in analytic terms, while the latter can be
computed via N-body simulations or by semi-analytic methods
\cite{NMCD} (since the velocity 
dispersion of galaxies in a cluster reflects the cluster's potential
well depth, a directly predictable quantity in spherical-collapse models).   Cluster 
counts as a function of $\sigma$ and $z$ are a more sensitive
probe of DE than is the abundance as a function of $z$
alone. 

\begin{figure}
\centering
\epsfig{width=4in, angle=90, file=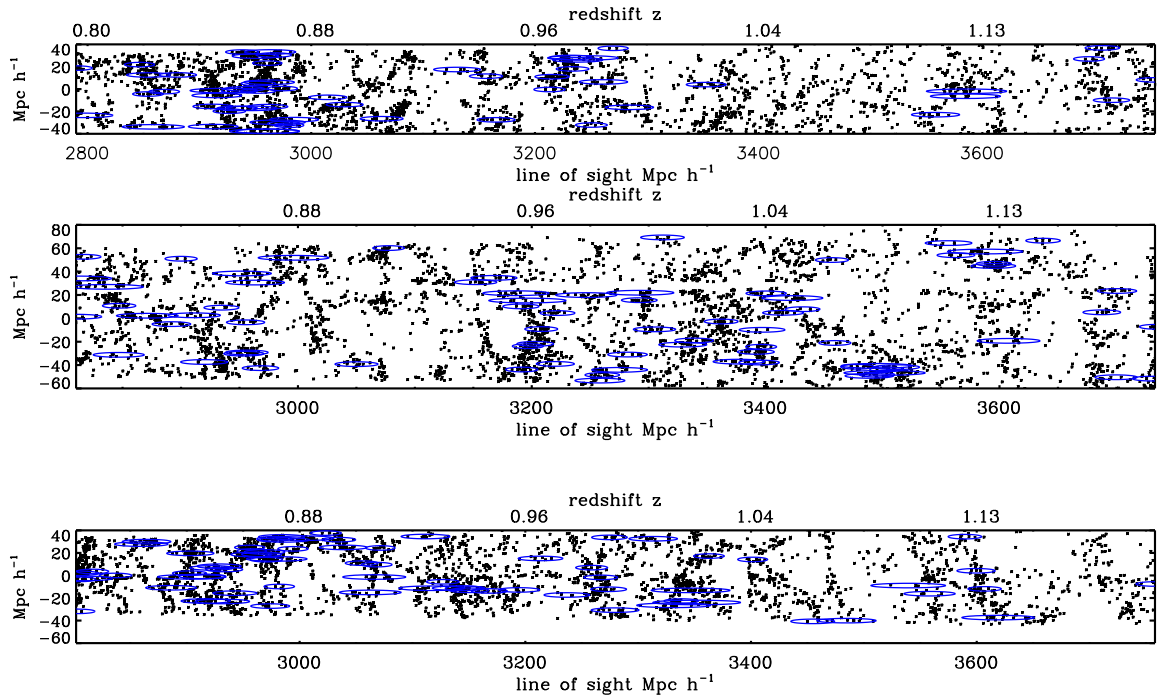}
\caption{Galaxies, groups and clusters in the DEEP2 Redshift
Survey.  The positions of galaxies in redshift space are indicated by
points, while ellipses show the positions of groups and clusters.  The
major axes of the ellipses are proportional to the group and cluster
velocity dispersions.}
\label{fig:clust_z}
\end{figure}

Groups and clusters of galaxies in the DEEP2 survey are detected with 
an automated cluster-finding algorithm that searches for overdensities
of galaxies in redshift space \cite{Marinoni, Gerke}.  Importantly,
this algorithm detects clusters regardless of the properties of their
member galaxies, so it does not require knowledge of the evolution of
those properties.  We have already applied this algorithm to
the DEEP2 data; the positions of some DEEP2 groups and clusters are
shown in Figure~\ref{fig:clust_z}.  

Since this cosmological test relies on the \emph{evolution}
of cluster abundance, it is useful to have a local sample of groups and
clusters against which to compare the $z\sim 1$ sample detected in
DEEP2.  Existing data from 2dF and SDSS should be sufficient for this
purpose.  

\section{Necessary Precursors}
\label{sec:precursors}

The crucial DEEP2 and SDSS data for this test are already in hand; no further
observations are required. The developments needed now
lie in the realms of simulation and data analysis.  

First, it is
necessary that we understand how the velocity dispersion of \emph{galaxies},
which we can measure, relates to the velocity disperison of \emph{dark
matter}, which is what we can most easily predict.  There is reason to
believe that the 
two are not equal---i.e., that there is a so-called ``velocity
bias'' in galaxy clusters.  Attempts are
underway to model this 
using N-body and hydrodynamic simulations \cite{Colin, Diemand, Gao};
if it is not constrained, 
the velocity bias could be a significant source of systematic error in
the DEEP2 cosmological constraints.

Second, because constraints on $w$ are strongly degenerate with other
cosmological parameters that are measurable at low redshift, the
power of DEEP2 to constrain DE would be 
greatly improved if these parameters could be fixed elsewhere.  For
example, it is expected that the Sloan Digital Sky Survey (SDSS) will
be able to constrain $\sigma_8$ independent of galaxy cluster
searches; great progress is now being made on techniques for this
\cite{Seljak, Abazajian}.  Because the local cluster abundance
primarily depends on a degenerate combination of $\sigma_8$ and
$\Omega_m$, fixing $\sigma_8$ will then provide a tight constraint in
the $\Omega_m$ direction (with modest $w$ degeneracy) from SDSS
clusters.  The high-redshift cluster abundance depends sensitively on
these two parameters; if they are poorly known, the constraints on DE
available from DEEP2 will be greatly degraded.  In all parameter
constraint plots shown, we assume that $\sigma_8$ is known and that a
cluster sample has been obtained over the entire SDSS volume with
systematic errors equal to those in DEEP2; constraints would be
slightly different if both $\Omega_m$ and $\sigma_8$ were simply
fixed.   

Finally, looking to the future, we expect that a potential twentyfold
larger future survey modeled on DEEP2 would require a 
preliminary deep, wide-field photometric survey, covering tens of
square degrees in at least three photometric bands, to allow selection
of spectroscopic targets at high redshift.

\section{Error Budget}

\subsection{Statistical Errors}
\begin{figure}
\centering
\epsfig{width=5in, file=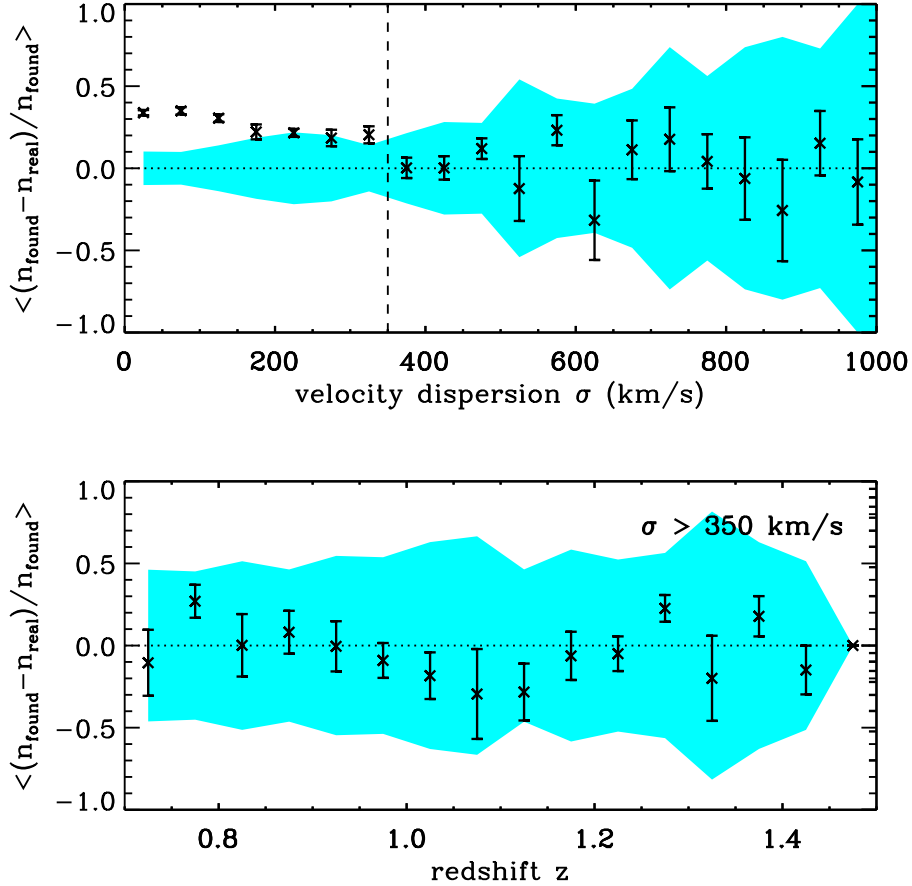}
\caption{Fractional errors in measuring $dN/d\sigma)$ and $dN/dz$.
Upper panel: the data points show the fractional systematic error
  $\langle\delta_N\rangle$
  as a function of velocity dispersion,
  estimated by running the DEEP2 cluster-finding algorithm on twelve
  independent mock DEEP2 pointings.  Error bars show the standard
  deviation of 
  the mean $\sigma_{\langle\delta\rangle}$, while the shaded region
  shows the  fractional cosmic variance 
  (plus Poisson noise) $\sigma_{cos}$ for a single DEEP2 field
  ($120\times 30$ arcmin---$30\%$ of the full area), in bins of
  $50\kms$.  For $\sigma \ge 350\kms$, the systematic errors are
  dominated by cosmic variance.   
  Bottom panel: Fractional systematic error and
  fractional cosmic variance as a function of redshift in bins of
  0.05 in $z$, after  
  groups with $\sigma < 350\kms$ have been discarded.  
  Systematic offsets are smaller than the cosmic variance.}
\label{fig:errs}
\end{figure}

There are two sources of statistical error in this
experiment---Poisson error and cosmic variance.  The former is just
the $\sqrt{N}$ uncertainty expected in counting $N$ objects, while the
latter is the excess variance in $N$ that arises
because galaxy clusters are themselves clustered together, rather
than being randomly distributed in space, so that no finite volume of
space constitutes a completely fair sample.  The fractional effect
of these two types of error on the measured cluster abundance is shown
by the 
shaded region in Figure~\ref{fig:errs}.  

\subsection{Systematic Errors}

In addition, there are three principal sources of systematic error.
The first 
is a measurement bias in $dN/d\sigma dz$ due to errors in
cluster detection.  Any automated cluster-finding algorithm in
redshift space is subject to some level of contamination from false
detections and incompleteness due to missed clusters.
Furthermore, the membership of individual clusters is often
imperfectly reconstructed, which leads to errors in measured velocity
dispersions.  These errors can lead to systematically incorrect
measurements of $dN/d\sigma dz$.  

To guard against this source of
error, we have calibrated our cluster-finding algorithm for DEEP2
by applying it to a set of twelve mock galaxy catalogs~\cite{YW} with
the same geometry as the DEEP2 survey.  These have been created by 
populating dark-matter-only N-body simulations with galaxies according
to the so-called ``halo model.'' This model places galaxies in dark
matter halos according to a halo-occupation distribution (HOD), which
specifies the average number of galaxies occupying a halo of mass $M$.
As shown in Figure~\ref{fig:errs}, when the cluster finder has been
properly calibrated, measurement bias errors are substantially smaller
than the 
expected statistical error in the DEEP2 sample for velocity
dispersions $\sigma \ge 350$ km/s. 

This calibration method leads to a second potential source of
systematic error, however.  The HOD used to create the DEEP2 mock
catalogs is chosen to be consistent with the two-point correlation
function and number density of DEEP2 galaxies, but it is not uniquely
specified by these statistics.  If the HOD we use for calibration does
not perfectly reflect the real universe, it is possible that our
cluster-finder will be mis-calibrated.  We have explored the
systematic errors introduced by changing the HOD, and we find that,
for HODs that are consistent with DEEP2, such errors are substantially smaller than
the expected statistical error.

The final major potential source of systematic error in this experiment is
the velocity bias discussed in Section~\ref{sec:precursors}.  This
parameter is defined as the ratio of the velocity dispersion of
galaxies in clusters to the velocity dispersion of the dark matter
particles: $b_v=\sigma_{gal}/\sigma_{dm}$.  There is evidence from
simulations \cite{Colin, Diemand, Gao} that $b_v$ differs from unity
by $\sim 10\%$, but so far there is debate about the 
magnitude of this effect. Because $dN/d\sigma$ varies rapidly with
$\sigma$, a $10\%$ systematic error in $\sigma$ could translate to a
much larger ($\sim 40\%$) error in the abundance.  Hence, unless the
value of $b_v$ can be better constrained by simulations, it will be a
dominant source of uncertainty in this experiment.  Fortunately, as
N-body and hydrodynamical simulations improve, it
should be possible to determine the value of $b_v$ separately from measurements of $w$.

Finally, it is important to note how the error budget would change in
the case of a significantly larger survey.  If it were possible to
cover an area 20 times larger than DEEP2, the cosmic variance would be
reduced by a factor of $\sim \sqrt{20}$ (in the most pessimistic scenario in which a single large field is surveyed, cosmic variance errors will only will be reduced by $\sqrt{13}$).   In such a scenario, the systematic errors discussed above would dominate the measurement.  A large survey would therefore necessitate a more accurate cluster-finding method, as well as a more
well-constrained HOD and velocity bias parameter.

\subsection{Assumed Priors}

We assume two priors in parameter space when we estimate
parameter constraints that will be possible with DEEP2.  We
first assume that the universe is flat, i.e., that
$\Omega_\Lambda=1-\Omega_M$ at late times.  We further assume that
SDSS will fix the value of $\sigma_8$ to high precision (i.e., well
enough that
uncertainty in $\sigma_8$ is subdominant to the other uncertainties in
the measurement; roughly, a 5\% error in $\sigma_8$ leads to a 10\% error in $w$ in this method). 

\section{Expected Constraints on Dark Energy}

\begin{figure}
\centering
\epsfig{width=5.in, file=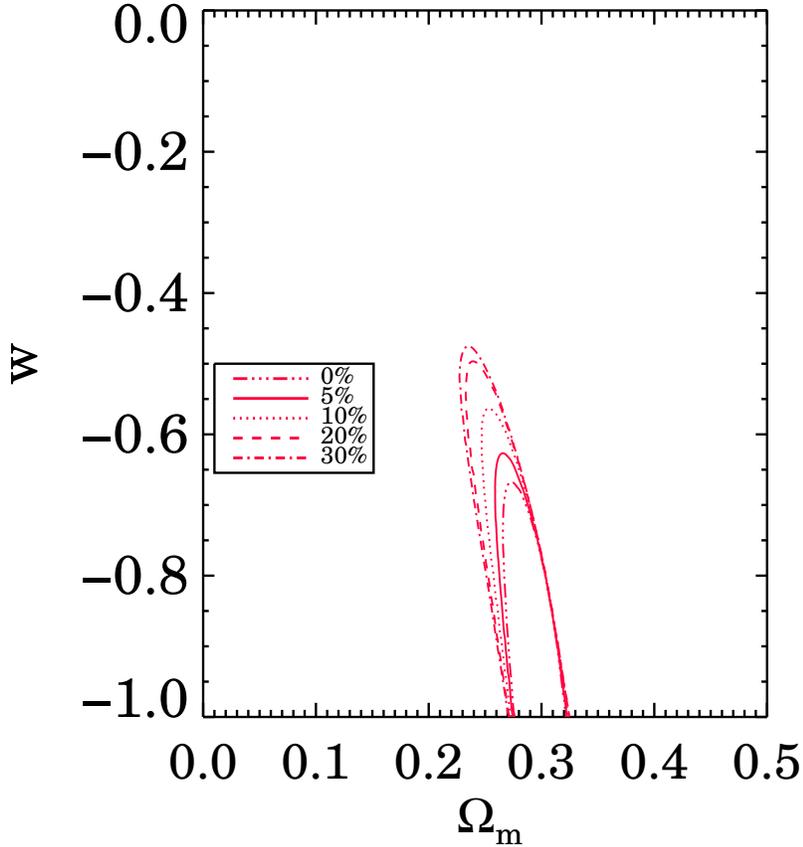}
\caption{Cosmological constraints expected from measuring the group
cluster abundance as a function of $\sigma$ and $z$ in the DEEP2
Galaxy Redshift Survey.  Contours show projected $95\%$ ($2\sigma$)
confidence regions for a fiducial cosmology with $w=-1$ and
$\Omega_M=0.3$ (the ``vanilla'' $\Lambda$CDM model).  Shown also are
projected contours including various levels of
systematic error in the measured abundance (fully covariant
amongst all redshifts and velocities).  All contours assume that
the velocity function $dN/d\sigma$ can be accurately measured down to
$350 \kms$. As shown, DEEP2 will be
able to constrain $w$ to $\sim 20\%$ ($1 \sigma$) if systematic errors can be
controlled to better than the $10\%$ level.} 
\label{fig:cont_minus1}
\end{figure}

\begin{figure}
\centering
\epsfig{width=5.in, file=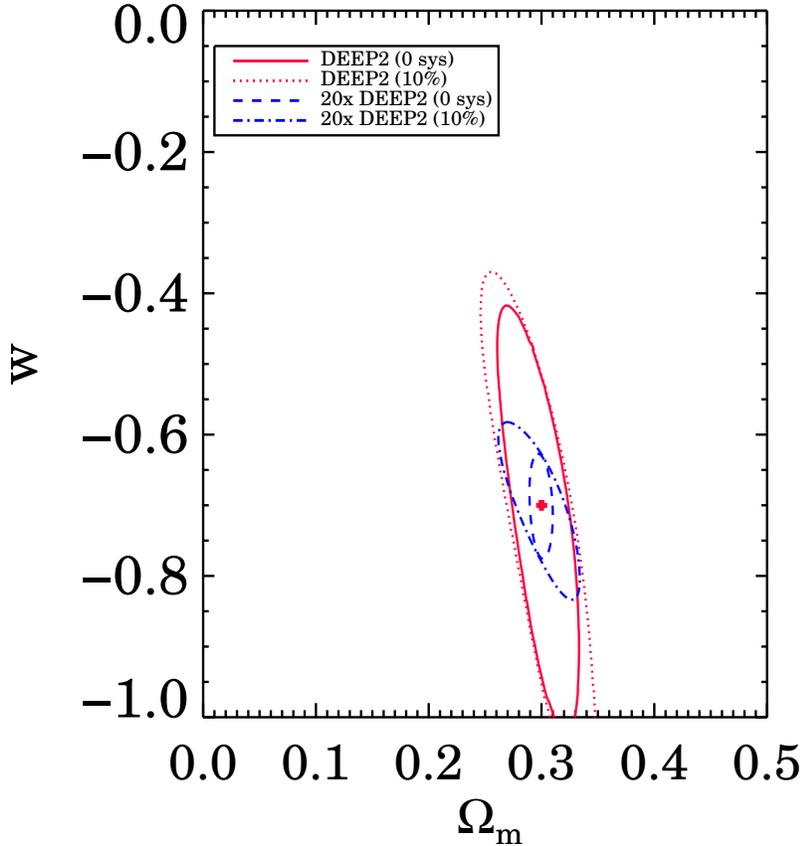}
\caption{Constraints expected from DEEP2 and from a proposed survey
covering 20 times as much area as DEEP2.  As in
Fig.~\ref{fig:cont_minus1}, contours show $95\%$ confidence regions,
with various levels of systematic error, but in this case the fiducial
cosmology has a constant DE equation of state $w=-0.7$.  In this scenario,
DEEP2 will only achieve a marginal rejection of the ``vanilla'' $w=-1$
case, whereas the larger survey will achieve a very significant
rejection, even without inclusion of complementary constraints from other methods.  }
\label{fig:cont_07}
\end{figure}

The constraints that will be possible by combining measurements of the cluster velocity function from DEEP2 and SDSS are shown in
Figure~\ref{fig:cont_minus1}.  In this figure, we show the effect of various levels of systematic uncertainty on the strength of the test, based on the assumption that systematics are completely covariant amongst all redshift and velocity bins (i.e., the most pessimistic sort of systematic effect possible).  As shown, if systematic errors can be
controlled well (to the $<10$ percent level), then DEEP2 can
constrain $w$ to $\sim 20\%$ ($1 \sigma$).   For a survey covering $\sim 20$
times the area of DEEP2, these constraints can be improved by a factor
of {$>2$} \emph{if} systematic effects can be controlled well
enough that statistical errors dominate \emph{and} appropriate prior
constraints on $\sigma_8$ are available.  As shown in
Figure~\ref{fig:cont_07}, DEEP2 will not be sufficient to
distinguish a cosmology with $w=-0.7$ from a $\Lambda$CDM model with
high conficence, but the larger survey would easily distinguish
the two.  We have not projected constraints on models with varying $w$
here because the main subject of this paper, DEEP2, has no hope of
constraining such models.  In these figures, we have used the absolute abundance of groups at low redshift and high redshift as separate constraints.  If we instead use the \emph{ratio} of abundance at high $z$ to low $z$, along with the absolute abundance at low redshift, to measure cosmological parameters we get constraints of very similar strength, but slightly different orientation in the $\Omega_m -- w$ plane.  This ratio method can potentially reduce the impact of systematics if they are similar at low and high redshift. 

\section{Risks and Strengths}

The principal risks for the cluster-abundance method relate to the
improved theoretical understanding of galaxy formation in clusters
that is required.  As mentioned above, uncertainty in the velocity
bias $b_v$ will be a dominant source of systematic error unless
simulations can constrain its value well.  Furthermore, uncertainty in
the HOD may be a problem for a larger survey with smaller statistical
errors.  Fortunately, both of these issues are areas of active
theoretical study and observational constraints on each are improving
quickly, so it is reasonable to suppose that the situation will
improve in the near future. 

The most obvious strength of the method proposed here is that the
necessary data are already in hand from the DEEP2 survey.
An additional manifest strength is the existence of supplementary data
that will allow cross-calibration of the DEEP2 data and other methods (e.g., by comparing
velocity dispersions to X-ray temperatures and S-Z decrements).  More
generally, most algorithms for detecting groups and clusters in
spectroscopic surveys make 
no assumptions about the properties of cluster galaxies.
Redshift-space cluster searches thereby
avoid a potential pitfall of some other optical cluster-finding
methods (e.g., the Red-Sequence 
method \cite{Gladders}) that assume the properties of the cluster 
galaxy population will remain stable over many gigayears of cosmic
time, to $z>1$.  

A final strength of this method is the fact that it comes as a free
byproduct of any large future spectroscopic galaxy survey at $z~1$,
provided that the survey has sufficiently high sampling density
($\sim$ 25--50\% or better)
and spectral resolution ($R\sim 2000$ or higher),   
and that its targeting strategy is uniform with regard to galaxy
type. For example, if the proposed KAOS spectrograph undertakes a
survey intended to constrain cosmology with baryon 
oscillations in the matter power spectrum, it can simultaneously
measure cluster abundance evolution if the survey is properly
designed.    

\section{Future Needs}

Apart from the theoretical advances discussed above, no further
developments are needed to carry out the proposed test with DEEP2
data; the survey is already nearing completion, and large amounts of
SDSS data are in hand.  In order to expand this test to a larger
survey, it would be necessary to design and 
build a new multi-object spectrograph to be placed on a new or
existing 8-meter (or larger) telescope.  Moderate-to-high spectral
resolution ($R\ge 2000$) is required to resolve cluster velocity
structure, and the 
ability to observe hundreds or thousands of galaxies simultaneously is also
necessary for efficiency.  Finally, a significant
investment of telescope time would be required, with many hundreds of
nights' observation being necessary to complete such a survey.

\section{Project Timeline}

With data already in hand, initial constraints on $w$ are expected
from DEEP2 within a year.  Expanding this project to a significantly
larger survey would require at least a decade of planning,
instrument-building, analysis and observation, but could be combined
with other projects.

\newpage

\end{document}